
\input /home/moore/gould/maggiemac
\input /home/moore/gould/phyzzx
\hoffset=0.375in
\input /home/moore/gould/maggieref
\twelvepoint
\font\bigfont=cmr17

\bigskip
\centerline{   }
\vskip10mm
\centerline{\bigfont The Orientation of Spin Vectors}
\centerline{\bigfont of Galaxies in the Ursa Major Filament}
\smallskip
\centerline{\bf Cheongho Han}
\smallskip
\centerline{\bf Andrew Gould}
\smallskip
\centerline{Dept of Astronomy, Ohio State University}
\centerline{Columbus, OH 43210}
\vskip5mm
\centerline{\bf and}
\vskip5mm
\centerline{\bf Penny D. Sackett}
\smallskip
\centerline{Institute for Advanced Study, Princeton, NJ 08540}
\vskip6mm
\centerline{E-mail cheongho@payne.mps.ohio-state.edu}
\centerline{E-mail Gould@payne.mps.ohio-state.edu}
\centerline{E-mail psackett@guinness.ias.edu}
\bigskip
\centerline{\bf Abstract}


     Spin vectors of 60 galaxies in the ``Ursa Major Filament'' are obtained
from CCD images and spectroscopic determinations of the directions of rotation.
These data are used to remove the four-fold degeneracy introduced by the
projected
images of galaxies, making possible the complete reconstruction
of the galaxy spin vectors.
Several possible organizations of spin vectors are investigated by means of
statistical analyses of the three-dimensional spin vector maps.
The results indicate that there exists no significant
alignment of galaxy spin vectors with respect to the supergalactic plane,
the supergalactic axis, or the cylindrical axis of the filament itself.
No tendency for linear alignment of spin vectors towards any specific direction
was detected.
We also obtain a statistically negative result for azimuthal, radial, and
circular-
helical
orientation of spin vectors.

\endpage

\chapter{Introduction}

     The distribution of galaxy orientations within various clusters and the
Local
Supercluster (LSC) has been investigated by many researchers.
Rood {\rm et al.} (1972) investigated galaxy orientations in Coma Cluster,
and their pioneering work was followed by ellipticity studies of
Schipper \& King (1978) and Djorgovski (1983).
Anistropic distribution of galaxy orientations in the A999 and A2179
clusters have been reported by Strom \& Strom (1978).
Whereas Kapranidis \& Sullivan (1983) found no positive tendency of galaxy
alignmen
t in
the LSC, Jaaniste \& Saar (1978) stated that galaxy rotation axes tend to be
distributed parallel to the LSC plane.  Jaaniste \& Saar's study was criticized
and

reanalyzed by Flin {\rm et al.} (1983), who concluded that the projections
of galaxy rotation axes have a weak tendency to align toward the Virgo Cluster
cent
er.
Kashikawa \& Okamura (1992) examined the position angles and axis ratios of
618 galaxies in the Local Supercluster and concluded that there was a weak
tendency for spins of galaxies in the supergalactic plane to be
aligned with the plane and those away from the plane to be aligned
perpendicular to the plane.
Additional studies of galaxy orientations were made
by MacGillivray {\rm et al.} (1982), Binggeli (1982), Hawley \& Peebles (1975),
and Travese \& Cirimele (1992).

     Most previous studies have focused on the galaxy distributions in
clusters and most of them are based on two-dimensional data of position angles,
although three-dimensional data
have been used by Helou \& Salpeter (1982) and Hoffman {\rm et al.} (1989).
However, in these latter studies, the only pattern that was searched for was
a net linear alignment and the samples were restricted to  Virgo
cluster galaxies.
The distribution of galaxy orientations based on three-dimensional data for
other types of large-scale structures such as filaments and walls has not yet
been
investigated.

     Most previous studies have focused on the galaxy distributions in
clusters and most of them are based on two-dimensional data of position angles,
although three-dimensional data
have been used by Helou \& Salpeter (1982) and Hoffman {\rm et al.} (1989).
However, in these latter studies, the only pattern that was searched for was
a net linear alignment and the samples were restricted to  Virgo
cluster galaxies.
The distribution of galaxy orientations based on three-dimensional data for
other types of large-scale structures such as filaments and walls has not yet
been
investigated.

     Only in the last decade with the advent of large
three-dimensional (redshift) maps has the
existence of filaments become firmly established.
The standard theory of large-scale structure formation by gravitational
collapse is generally believed to account for the observed filamentary
structures.
Within the simplest version of this theory (see \S\ 7), spin vectors of
galaxies mi
ght
be expected to be randomly oriented.  Hence, the detection of any organization
of spin vectors would indicate either that the simplest version of
the standard theory is inadequate or perhaps that the entire standard picture
is wrong.
The purpose of this paper is to determine whether the spin vectors of
spiral galaxies in one filament, the ``Ursa Major Filament'' are aligned
or have any other form of recognizable organization.

\chapter{``Ursa Major Filament''}

     The ``Ursa Major Filament'' is an elongated
region of redshift space extending from approximately
$(-70,320,0)$ to $(800,1250,0)$ in the Supergalactic Redshift
Cartesian coordinates (SGC), which is overdense by more than a factor 2
compared to neighboring regions.
The supergalactic coordinate system (SGL,SGB) has its origin and north pole at
$l = 137^{\circ}\hskip-2pt.37, b = 0^{\circ}$, and
$l = 47^{\circ}\hskip-2pt.37, b = + 6^{\circ}\hskip-2pt.32$,
respectively.
In the SGC, SGX represents
the distance of a galaxy parallel to one of the Cartesian axes
in Supergalactic coordinates in ${\rm km}\,{\rm s}^{-1}$.
The SGX axis is directed towards the origin of the Supergalactic
coordinates, ${\rm SGL} = 0^{\circ}$, ${\rm SGB} = 0^{\circ}$.
The SGZ axis is perpendicular to the Supergalactic plane, towards
${\rm SGL} = 90^{\circ}$, ${\rm SGB} = 90^{\circ}$ (Tully 1987).

     The galaxies in the Ursa Major Filament have redshifts between
approximately 500 ${\rm km}\,{\rm s}^{-1}$ and 1300 ${\rm km}\,{\rm s}^{-1}$
and lie between $l\sim 210^{\circ}, b\sim 85^{\circ}$ and
$l\sim 140^{\circ}, b\sim 60^{\circ}$ in galactic coordinates.
Most galaxies are located around supergalactic XY plane; $\sim$
200 ${\rm km}\,{\rm s}^{-1}$ below and above the supergalactic plane.
The axis of cylindrical symmetry of the Ursa Major Filament lies
in the supergalactic plane and makes a $\sim 47^{\circ}$
angle with SGX.
Note that the name ``Ursa Major Filament'' is our invention.

\chapter{Observations and Data Reduction}

     Our full sample consists of all spiral galaxies with redshifts listed
in de Vaucouleurs {\rm et al.} (1991, hereafter RC3)
that lie within a cylindrical tube in redshift space that is
400 ${\rm km\,s^{-1}}$ in diameter and 1275 ${\rm km}\,{\rm s}^{-1}$ in length.
The end points of the cylinder lie at $(-70,320,0)$ and $(800,1250,0)$ in SGC.
Spirals are defined here as galaxies having a type 1-8 as listed in RC3.
The positions of galaxies in redshift space are determined by correcting the
heliocentric velocity to the Local Group frame defined as motion toward
$\alpha = 22.32^{\circ}, \delta = 48.5^{\circ},$ at $300$ ${\rm km}\,{\rm
s}^{-1}$.
We attempted to obtain images and spectra of all 94 galaxies in the sample.
However, due to bad weather data were obtained for only 71 and 69 galaxies
respectively.

     CCD images in $V$ and $I$ bands were obtained for 71 galaxies in the
sample
on the nights of 15-22 April 1994 using the Ohio State University
Imaging Fabry-Perot Spectrograph (IFPS) on the 1.8m Perkins Telescope
of the Ohio State and Ohio Wesleyan Universities at Lowell Observatory.
We used the IFPS as a direct imaging focal reducing camera (no etalon in the
beam)
with the Lowell NSF TI 800$\times$800 CCD detector.
This CCD has 15$\mu$m pixels giving an image scale of $0''\hskip-2pt .49$
pixel$^{-
1}$.
Typical exposure times were 5-8 minutes in the V band, and 1-2 minutes in the I
ban
d
depending on the galaxy surface brightness.

     Long-slit spectra of 69 galaxies were also obtained using the
CCDS, an Ohio State Boller \& Chivens Spectrograph with a tektronix
$512 \times 512$ CCD detector, with 350 lines\ mm$^{-1}$ grating
in the second order  resulting in a resolution of 1.44{\rm \AA}\ pixel$^{-1}$.
The wavelength range is approximately 6200${\rm \AA}$-6700${\rm \AA}$.
The exposure times were typically 5-8 minutes per spectrum.
We relied primarily on the prominent ${\rm H\alpha} \lambda 6563$ emission line
to measure the rotation, but also made supplementary use of
$[{\rm NII}]\lambda\lambda 6548.1, 6583.4$, and $[{\rm SII}]
\lambda\lambda 6717, 6731$ lines.

     In order to determine which side of the galaxy is receding we placed the
long slit of the CCDS approximately along the major axis of the galaxy.
Slight misalignments of the slit relative to the major axis do not
affect this determination.
Thus, in order to minimize the need to rotate the slit, the measured
position angles (PAs) were classified into 6 groups: the first group contains
galaxies with PA in the range between $0^\circ$ and $29^\circ$, the 2nd group
in the range $30^\circ$ to $59^\circ$, and so on.
We then placed the slit at the central angle during the spectroscopy
of galaxies in each group.
Because the ${\rm H\alpha}$ line is identified without ambiguity, wavelength
calibration was not necessary in the data reduction.
Bias substraction and sky substraction of the spectra and the images
were carried out using the interactive data reduction package IRAF.
When necessary, we removed cosmic rays from the spectra using
software that we developed ourselves.

\chapter{Data Analysis}

     By combining the images and spectra that we obtained with archival data
from the RC3 and the Palomar Observatory Sky Survey (POSS) prints,
we attempted to establish 5 pieces of information for each galaxy:
the PA of the major axis,
the apparent axis ratio, the shape of the spiral arms (`N'-shaped
or `S'-shaped), the receding side of the disk, and
the side of the bulge blocked by a dust lane.
Our results are listed in Table 1.
The names, right ascension (RA), and declination (DEC) are listed in the
first 3 columns.
Column 4 gives the morphological type as listed in RC3.
The axis ratios, line-of-sight redshift velocities, and PA, in columns 5
through 7,
were obtained primarily from the RC3.
For some nearly face-on galaxies, the RC3 does not list a PA;
in these cases we measured the PA from V-I color maps constructed from the
IFPS images.

     The images were used primarily to determine the direction of rotation
of the spiral arms.
When possible, we also noted the side of the bulge blocked by dust.
The angle of the dust relative to the central bulge is
listed in column 10 of Table 1: the angle is measured from the north to the
east.
For example, $135^{\circ}$ means that the south-east side of the
bulge is blocked by dust and hence is closest.
When we could establish these two bits of information,
it was usually from the V image alone.
However, we double-checked uncertain determinations by inspecting a
V-I color map of the galaxy.
The spiral arm pattern  of each galaxy is listed in column 8 of Table 1.
When the spiral pattern appears S-shaped, it is marked with `S'.
The opposite pattern is marked with `N'.
Column 9 is a ``clearness code'' for our spiral pattern determination.
`A' means that the spiral arm structure is discernable without difficulties,
while `E' means that it is impossible to discern.
Between `A' and `E', the clearness code is divided into three classes.
Galaxies with uncertain but discernable structure are classified as `D1'.
Galaxies with an ambiguous structure on which we unanimously agree are assigned
`D2
'.
Finally, galaxies with some pattern but without unanimous agreement are
classified
as `D3'.
In our analysis we do not include galaxies which are designated as `D3'.
Edge-on galaxies are classified as either `B' or `C': `B' with visible dust
lane
and `C' without detected dust lane.
The prefix `P-' means that the pattern is identified from the POSS prints
rather than the IFPS data.

     The last two columns in Table 1 give information about the receding side
of
the disk as determined from the long-slit data.
The penultimate column is marked `E' or `W' according to whether the receding
side
is more closely aligned with east or west.
Since the slit positions are restricted to 15$^{\circ}$, 45$^{\circ}$,...,
165$^{\c
irc}$,
there is no ambiguity in the designation of receding side.
The final column is a ``line clearness code'' for the receding side
determination.
A `1' means that the receding side can be identified
from ${\rm H}\alpha$ by eye without ambiguity.
When the ${\rm H}\alpha$ trend is not obvious, we use software that can detect
smal
l
differences in radial velocity.
The program is designed to average the flux of the ${\rm H\alpha}$ line from
both
sides of the slit and plots average flux with respect to the wavelenth.
If the receding side is obviously resolved by this program,
we classify the clearness as `2'.
A `3' designates that some ambiguity remains after using the program.
Lines which are more ambiguous  but not impossible to discern are classified as
`4'